\newif\ifpreprint

\preprinttrue

\ifpreprint
    \documentclass[utf8]{FrontiersinHarvardPreprint}
\else
    \documentclass[utf8]{FrontiersinHarvard}
\fi

\usepackage{url,hyperref,microtype,subcaption}
\usepackage[onehalfspacing]{setspace}
\usepackage{lineno}

\ifpreprint
\else
    \linenumbers
\fi

\usepackage[utf8]{inputenc} 
\usepackage[T1]{fontenc}    
\usepackage{url}            
\usepackage{booktabs}       
\usepackage{amsmath,amssymb,amsthm}
\usepackage{amsfonts}       
\usepackage{nicefrac}       
\usepackage{makecell}
\usepackage{microtype}      
\usepackage{mathrsfs}
\usepackage{multicol}
\usepackage{graphicx}
\usepackage{doi}
\usepackage{float}
\usepackage{acronym}
\usepackage{listings}
\usepackage[obeyclassoptions]{standalone}
\usepackage{tikz}
\usepackage[dvipsnames]{xcolor}

\newacro{abm}[ABM]{Agent-Based Model}
\newacro{ca}[CA]{Cellular Automaton}
\newacro{ib}[IB]{Individual-Based}
\newacro{ode}[ODE]{Ordinary Differential Equation}
\newacro{pde}[PDE]{Partial Differential Equation}
\newacro{dnn}[DNN]{Deep Neural Network}
\newacro{mbs}[MBS]{Many-Body Simulation}
\newacroplural{ca}[CA]{Cellular Automata}
\newacro{gpu}[GPU]{Graphical Processing Unit}

\definecolor{COLOR1}{HTML}{6bd2db}
\definecolor{COLOR2}{HTML}{0ea7b5}
\definecolor{COLOR3}{HTML}{0c457d}
\definecolor{COLOR4}{HTML}{ffbe4f}
\definecolor{COLOR5}{HTML}{e8702a}

\def\keyFont{\fontsize{8}{11}\helveticabold }
\def\firstAuthorLast{Pleyer {et~al.}} 
\def\Authors{Jonas Pleyer\,$^{1,*}$
\ifpreprint
\else
, Christian Fleck\,$^{1}$
\fi}
\def\Address{$^{1}$Freiburg Center for Data Analysis, Modeling and AI, University of Freiburg, Freiburg im Breisgau, Germany}
\def\corrAuthor{Jonas Pleyer}

\def\corrEmail{jonas.pleyer@fdm.uni-freiburg.de}

\author[\firstAuthorLast ]{\Authors} 
\address{} 
\correspondance{} 

\extraAuth{}

\begin{document}
\onecolumn
\firstpage{1}

\title[]{Agent-Based Modelling in Cellular Biology\\Are we flexible yet?}

\author[\firstAuthorLast]{\Authors}
\address{}
\correspondance{}

\maketitle


\begin{abstract}
    \section{}
    Cellular \aclp{abm} are commonly employed to describe a variety biological systems.
    Over the course of the past years, many modeling tools have emerged which solve particular
    research questions.
    In this short opinion piece, we argue that existing frameworks lack flexibility compared to the
    inherent underlying complexity that they should be able to represent.
    We extract overarching principles of widely used software solutions across multiple domains and
    compare these with existing \acp{abm}.
    We come to the conclusion that existing \acp{abm} lack in flexibility which hinders overall
    progress of the field.

    \tiny
     \keyFont{ \section{Keywords:} Agent-Based Model, Individual, Cell, Biology, Library, Flexibility} 
\end{abstract}


\section{Introduction}
\acp{abm} have become indispensable tools in the study of complex systems.
Their applications cover topics in Ecology~\cite{Grimm2013}, Social Sciences~\cite{Bankes2002},
Autonomous Cars~\cite{Karolemeas2024}, Spread of COVID infections~\cite{Shattock2022} and many more.
They utilize descriptions on the level of individual entities such as cells, organisms or humans,
called agents, to build up models of complex systems.
They aim to study the emergence of collective behaviors as a result of \ac{ib} interactions.
In the field of biology, cells can be considered the fundamental building blocks of nature,
comprising many complex systems such as bacterial communities~\cite{Nagarajan2022},
organs~\cite{DuttaMoscato2014}, plants~\cite{Merks2011} and tissues~\cite{Thorne2007}.
They are particularly suited to capture effects such as heterogeneity, spatial phenomena and provide
a natural way to think in terms of cellular processes.

In order to quickly construct new simulations, many frameworks have emerged which simplify
the process of model design~\cite{Pleyer2023}.
However, despite their wide applicability, most of these models rely on a cellular representation
which is baked into the respective \ac{abm} and can not simply be modified by the user.
Furthermore, these models often come with a large set of parameters that need to be specified in
order to obtain a working simulation which results in problems with respect to interpretability and
parameter estimation of the model.

With this text, we argue that cellular \acp{abm} need to become more flexible with respect to design
of agents and environment in order to tackle the aforementioned challenges.
We further argue that many groundbreaking modeling techniques have been preceded by strong
results and made popular by generalist tools which are able to solve a whole class of problems,
thereby enabling many researchers to access the novel method.

\section{Principles of Foundational Tools}

Over the past years, a variety of foundational techniques have greatly influenced scientific
research by enabling advanced workflows and providing novel insights.
We focus on a subset of methods which are so fundamental that their usage spans across almost all
disciplines and can target a variety of problems.
Due to their foundational nature, each method comes with many tools which have been developed such
that researchers are not concerned with implementation details but can focus on the scientific
question at hand.

\subsection{Selected Foundational Methods}
\acp{ode} are one example of a fundamental method which is applied across all sciencies and are
mostly used to describe the dynamics of various systems.
Their mathematical form is rather simple which makes it easy for computationally inclined
researchers to implement numerical solvers and has led to the developmend of many numerical
libraries which are frequently reused.
Examplse are the Boost library~\cite{Mulansky2011} or the Julia package
DifferentialEquations.jl~\cite{rackauckas2017differentialequations}.
When considering spatial effects, \acp{pde} provide a natural way to extend \acp{ode} although the
added structure also requries deeper mathematical knowledge and more complex numerical solvers.
Widely used tools are OpenFoam~\cite{Weller1998} for fluid simulations or
FEniCS~\cite{BarattaEtal2023,ScroggsEtal2022,BasixJoss,AlnaesEtal2014} for finite-element workflows.
In contrast, \acp{mbs} describe the dynamics of spatial systems on the basis of many interacting
particles.
GROMACS~\cite{Abraham2015} and LAMPPS~\cite{Thompson2022} both target classical molecular dynamics
where the latter focusses on materials modeling but also arguably provides better flexibility while
the latter is considered to be faster.
Finally, in the past years, \acp{dnn} have been applied to a variety of problems such as protein
folding~\cite{Jumper2021} or biomedical image segmentation~\cite{Ronneberger2015}.
These have been enabled by tools such as PyTorch~\cite{Ansel_PyTorch_2_Faster_2024} and
TensorFlow~\cite{tensorflow2015-whitepaper}.


\subsection{Shared Principles}

Despite their differences in applications and technical details, all of these tools exhibit similar
unifying characteristics.
Every individual component of these tools can be formulated using mathematical methods although
numerically obtained results are often not derivable by a purely mathematical approach.
This is also true for computations involving stochatic effects such as stochastic \acp{ode} or
approximate calculations.
This generalized concept can be observed across all levels of complexity.
Furthermore, since many of these methods have been established for numerous years, they follow
mathematical notions which are rooted in a description that is mostly agreed upon in the
respective field.
A particular problem is defined by a given a set of initially known properties (i.e. initial values,
labeled data, etc.) and a mathematical description.
In the case of dynamical systems one could ask what the temporal evolution of a given set of initial
values is over a given time interval.
But other questions such as: "Which parameters allow me to represent this set of datapoints most
optimally?" are also valid.
The mathematical description links these quantities together and a suitable solver can be chosen
which accomplishes the mapping most effectively.

The previously mentioned tools have only emerged once the corresponding method was already popularized.
However, the reason that these methods became popular in the first place was due to their strong
results either in academic or industrial settings.
Tools which make it easy to exploit these methods and deliver exceptional performance are a key
cornerstone for why these methods continued to be popular after initial success.
One prominent example where a technological advancement led to the popularization of a particular
method can be seen in the case of \acp{dnn}.
Training these networks on \acp{gpu}~\cite{Raina2009} via the gradient descent~\cite{Rumelhart1986}
optimization method was crucial for their breakthrough and only recently recognized with the nobel
prize~\cite{Werbos1982,Linnainmaa1976,Nobel2024Physics}.

Finally, another important aspect is that the presented methods have the capability to be very
complex but also to be arbitrarily reducable such that no parameters are left.
This means that in the case of an \ac{ode} given by
\begin{equation}
    \partial_t \textbf{x} = \textbf{f}(t,\textbf{x})
\end{equation}
any mathematically expressable term can be inserted for $\textbf{f}(t,\textbf{x})$, allowing for
complex models.
But despite this flexibility, we can still choose $\textbf{f}=0$, thus eliminating any model
properties.
These statements hold true for \acp{ode}, \acp{pde} and most \acp{mbs}.
It is a key feature that all of these tools allow scaling of the complexity scale without requiring
a given starting point which enables core modeling techniques such as model reduction and model
validation.

\begin{figure}[H]
    \centering
    \includegraphics[width=0.5\textwidth]{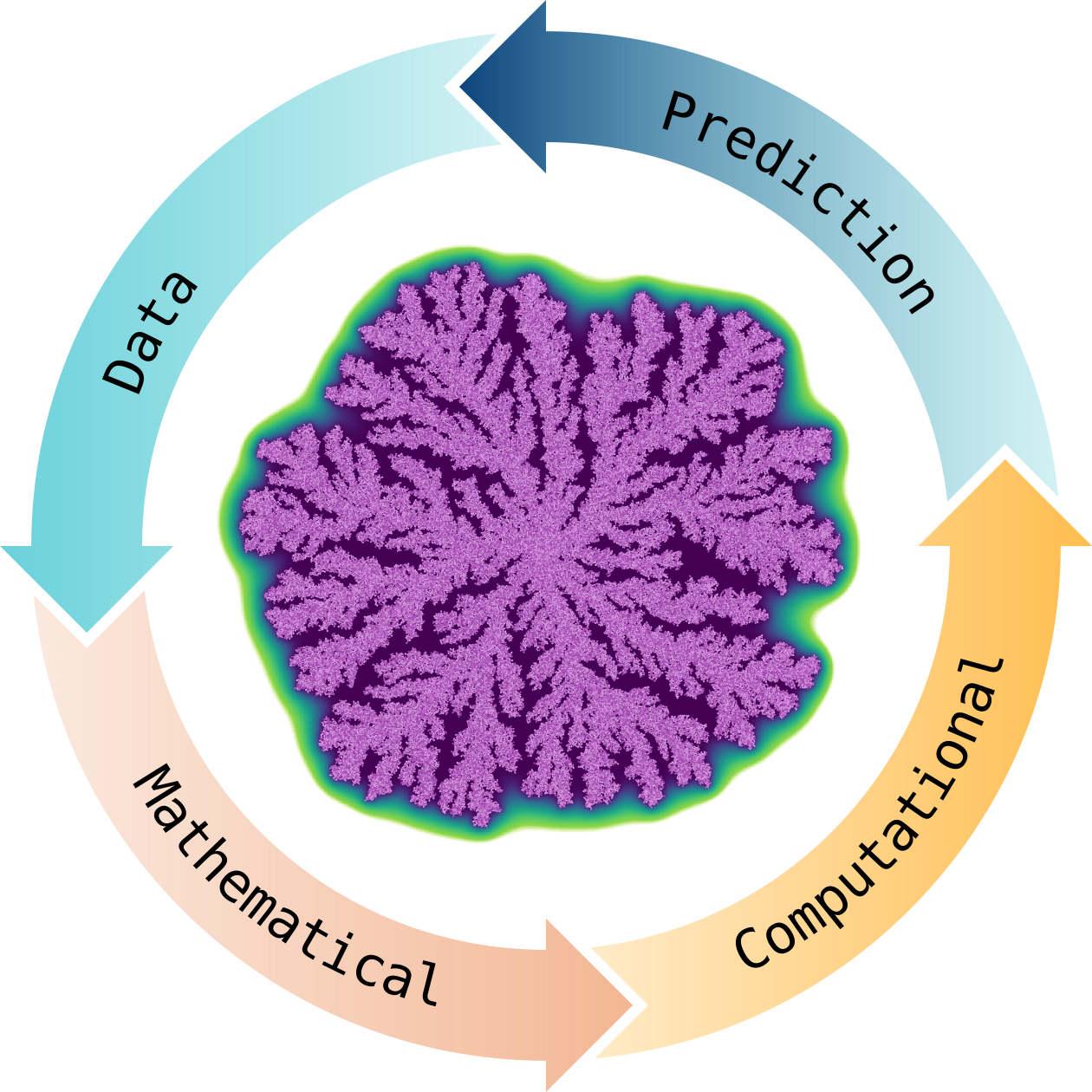}
    \caption{
        Schematic representation of a typical modeling approach.
        It is in principle possible to start at any arrow.
        A mathematical model is implemented within a computational framework which results in
        predictions that can be compared with biological data.
        This comparison nurishes our understanding of reality and allows us to formulate more
        appropriate models which starts the cycle again.
    }
    \label{fig:modeling-schematic}
\end{figure}

\subsection{General Modeling Approach}
In order to construct models which can describe biological systems and interactions therein, a common
workflow is usually applied.
This scheme (see Figure~\ref{fig:modeling-schematic}) involves multiple steps: A mathematical model
is translated into computational form which provides novel insights which can finally be compared
with data in order to embed the model within biological reality.
To close the circle, researchers have to align the mathematical model to the biological reality.
The model can be extended by including additional effects or simplified i.e. for parameter
estimation purposes.
Furthermore, when constructing new models it is important to be able to build up models from scratch
without having to rely on existing work such that minimal models with low amounts of parameters can
be constructed.

\section{Constructing Cellular Agent-Based Models}
When constructing an \ac{abm}, researchers are faced with the question if they should start
bottom-up, by defining core ingredients of the model by themselves or if a preexisting \ac{abm} can
be reused for their purpose.
Both approaches bear challenges and provide advantages which we want to highlight now.

\subsection{Using an existing \acs{abm}}
In our previous work, we have reviewed a variety of \acp{abm}~\cite{Pleyer2023}.
Almost all of them assume a particular cellular and environmental representation.
The most common choice for cellular shape is the soft-spheroid
model~\cite{Ghaffarizadeh2018,Gorochowski2012} while only a few select \acp{abm} providing support
for ellipses, cylinders~\cite{Kang2014} or other representations.
Although more variability exists for the description of extracellular compounds, most models choose
a diffusion-based approach~\cite{Breitwieser2021}.
Any particular choice for these representations is fundamental to the dynamics which these
models are capable of describing and will carry over in other aspects such as growth, physical
interactions and intracellular processes.\\
Furthermore, we observed that existing \acp{abm} carry a large set of non-optional parameters which
need to be specified in order to be able to use the model.
These parameters can be problems for methods such as model reductions since they are intrinsic to
the chosen \ac{abm} and can not be simply removed.
If they can not be known from literature values or determined by parameter estimation techniques,
researchers need to take great care in how the values of said parameters influence obtained results.
Before choosing a particular \ac{abm}, researchers need to evaluate if the model in question is
capable of describing the desired phenomena and can be fully parametrized by the given biological
context.
Once work has started within a particular modeling environment, switching to a different \ac{abm} is
often tedious and costly in time.
These observations present significant practical limitations on the ability of researchers to find a
suitable model for the given biological problem at hand.
They also lead to problems when employing model reduction techniques and make it difficult to follow
best principles such as Ockam's Razor~\cite{Sober2015} during model development.

\subsection{From a clean Slate}
Another option is to build up a model starting from a clean slate.
When deciding on this approach it is still possible to harness existing tools as long as they are
capable of describing the desired cellular properties and provide enough flexibility to perform
model reduction and parameter estimation techniques.
General-purpose tools such as Netlogo~\cite{Wilensky_NetLogo_1999} which are not specifically
tailored towards biological questions can thus be a viable option.
However, due to their generalized nature, they may lack desired functionalities specifically
taylored to biological problems and will thus involve additional work during model development.\\
The case where no existing framework is used and almost all code is written by the researchers
themselves, presents a very labor-intensive approach to solving the problem, even when utilizing
libraries for linear algebra and numerical solvers.
This approach also involves another class of challenges involving correctness of implemented
algorithms and accuracy of numerical results.\\
The resulting \ac{abm} is often highly tailored to a specific niche or the respective research
question at hand, thus limiting its ability to be reused in other contexts.

\subsection{Building Blocks}
In contrast to the two approaches described before, there exists another path which has not been
fully realized yet.
That is to disect the components of cell-agents into building blocks.
In order for this approach to be valuable, researchers need to be able to construct external
building blocks without having to rely on existing functionalities.
Furthermore we believe that widespread adoption can only be achieved if a tool can provide
sufficiently many predefined building blocks which can be reused by researchers for their respective
field of research.
This procedure is closely followed in the development of large software libraries where it is
desirable to provide reusable parts that can be freely combined.\\
Our own \ac{abm} cellular\_raza~\cite{Pleyer2025} aims to push further into this direction by
separating abstract mathematical concepts from numerical solvers and implemented building blocks.
However, it is too early to tell if this particular approach has the potential to become the
de-facto standard for constructing novel \acp{abm}.\\
Another promising framework is BioDynaMo~\cite{Breitwieser2021} which also considers building blocks
and provides a modular design.
Although it exhibits groundbreaking performance it is yet to be applied across a diverse range of topics and only supports a small set of building blocks at the time of writing.

\section{Conclusion}
\acp{abm} have already proven to be useful tools in describing cellular systems and their emergent
phenomena.
Furthermore, due to their individual-based treatment of cells, they provide an intuitive way
of bridging the gap from single-cell studies to collective phenomena.
However, in order for these tools to become widely applicable and easily reusable, a more flexible
approach which allows better reusability and customizability is required.
Frameworks which are too specialized or lack in performance will be unable to support the
exploratory nature that the diverse field of biology has to offer.
Furthermore, a more unified mathematical treatment would allow for the
construction of more generalized libraries that can be reused by researchers across topics ranging
from microbiology over human stem cells to plant cells.
The construction of either a widely used generalized library or a unified mathematical framework,
would provide the necessary abstractions to allow researchers to flexibly conceive and implement
novel \acp{abm}.
Only then can this field live up to the standards of explorability and flexibility that are common
in other areas of research.

\newpage
\bibliographystyle{FrontiersinHarvard}
\bibliography{references}

\end{document}